\documentclass[aps,preprint,amsmath,amssymb,showpacs]{revtex4}
\usepackage{graphicx}

\begin{document}
\title{Anomalous Quartic $WW\gamma\gamma$ and 
$ZZ\gamma\gamma$ Couplings
in $e\gamma$ Collision With Initial Beams and Final State Polarizations}

\author{S. Ata\u{g} }
\email[]{atag@science.ankara.edu.tr}
\affiliation{Department of Physics, Faculty of Sciences,
Ankara University, 06100 Tandogan, Ankara, Turkey}

\author{\.{I}. \c{S}ahin}
\email[]{isahin@science.ankara.edu.tr}
\affiliation{Department of Physics, Faculty of Sciences,
Ankara University, 06100 Tandogan, Ankara, Turkey}

\begin{abstract}
The constraints on the anomalous quartic $WW\gamma\gamma$ and
$ZZ\gamma\gamma$ gauge boson couplings are investigated through the 
processes $e\gamma\to W^{-}\gamma\nu_{e}$ and $e\gamma\to Z\gamma e$. 
Considering the longitudinal and
transverse polarization states of the final W or Z boson and 
incoming beam polarizations
 we find 95\% confidence level limits on the
anomalous coupling parameters $a_{0}$ and $a_{c}$ with an
integrated luminosity of 500 $fb^{-1}$ and $\sqrt{s}$=0.5, 1 TeV
energies. Assuming the $W^{+}W^{-}\gamma\gamma$
couplings are independent of the $ZZ\gamma\gamma$ couplings 
we show that the longitudinal polarization state of the
final gauge boson improves the sensitivity to anomalous couplings 
by a factor of 2-3 depending on energy and coupling. An extra 
enhancement in sensitivity by a factor of 1.3 comes from a set of 
initial beam polarizations.
\end{abstract}

\pacs{12.15.Ji, 12.15.-y, 12.60.Cn, 14.80.Cp}

\maketitle
\section{Introduction}
Many predictions of Standard Model (SM) of electroweak interactions
have been tested with a good accuracy in the recent experiments at
CERN $e^{+}e^{-}$ collider LEP and Fermilab Tevatron and the experimental
results confirms the $SU_{L}(2)\times U_{Y}(1)$ gauge structure of SM.
However, self-interactions of gauge bosons
have not been tested with a good accuracy
and their precision measurements are in the scope of future experiments.
Therefore precision measurements of these couplings in the future colliders
will be the crucial test of the structure of the SM.
Deviation of the couplings from the expected values would indicate
the existence of new physics beyond the SM. In this work we analyzed
genuinely quartic $WW\gamma\gamma$ and $ZZ\gamma\gamma$ 
couplings which do not induce new trilinear vertices.

In writing effective operators for genuinely quartic
couplings we employ the formalism of \cite{kuroda}. 
Imposing custodial $SU(2)_{Weak}$ symmetry and local
$U(1)_{em}$ symmetry and if we restrict ourselves to C and P conserving
interactions, the effective lagrangian for the $W^{+}W^{-}\gamma\gamma$
couplings is given by,
\begin{eqnarray}
L&&=L_{0}+L_{c}
\end{eqnarray}

\begin{eqnarray}
L_{0}&&=\frac{-\pi\alpha}{4\Lambda^{2}}a_{0}F_{\mu\nu}F^{\mu\nu}
W_{\alpha}^{(i)}W^{(i) \alpha}
\end{eqnarray}

\begin{eqnarray}
L_{c}&&=\frac{-\pi\alpha}{4\Lambda^{2}}a_{c}F_{\mu\alpha}F^{\mu\beta}
W^{(i) \alpha}W_{\beta}^{(i)}
\end{eqnarray}
where $W^{(i)}$ is the $SU(2)_{Weak}$ triplet and $F_{\mu\nu}$ is the
electromagnetic field strength. Effective lagrangians (2) and (3) also give
rise to anomalous $ZZ\gamma\gamma$ couplings. For sensitivity calculations
to anomalous couplings we set the new physics energy  scale 
$\Lambda$ to $M_{W}$. The vertex functions for
$W^{+}(k_{1}^{\mu})W^{-}(k_{2}^{\nu})\gamma(p_{1}^{\alpha})
\gamma(p_{2}^{\beta})$ generated from the effective lagrangians
(2) and (3) are given by

\begin{eqnarray}
i\frac{2\pi\alpha}{\Lambda^{2}}a_{0}g_{\mu\nu}\left[g_{\alpha\beta}
(p_{1}.p_{2})-p_{2 \alpha}p_{1 \beta}\right]
\end{eqnarray}
\begin{eqnarray}
i\frac{\pi\alpha}{2\Lambda^{2}}a_{c}\left[(p_{1}.p_{2})(g_{\mu\alpha}
g_{\nu\beta}+g_{\mu\beta}g_{\alpha\nu})+g_{\alpha\beta}(p_{1\mu}p_{2\nu}
+p_{2\mu}p_{1\nu})\right. \nonumber \\
\left. -p_{1\beta}(g_{\alpha\mu}p_{2\nu}+g_{\alpha\nu}p_{2\mu})
-p_{2\alpha}(g_{\beta\mu}p_{1\nu}+g_{\beta\nu}p_{1\mu})\right]
\end{eqnarray}
respectively. For a convention, we assume that all the momenta are
incoming to the vertex.
The anomalous $ZZ\gamma\gamma$ couplings are obtained by multiplying 
(4), (5) by $\frac{1}{\cos^{2}\theta_{W}}$ and making $W \to Z$.

Indirect information on quartic gauge boson interactions comes from 
the fact that they modify gauge boson two point functions at one 
loop \cite{brunstein}. The Measurements at low energy and the Z pole 
give constraints on these couplings to be smaller than 
$10^{-3}-10^{-1}$ depending on the couplings. 

Measurements at CERN $e^{+}e^{-}$ collider LEP2 provide present collider
limits on anomalous quartic $WW\gamma\gamma$ and $ZZ\gamma\gamma$
couplings. Recent
results from OPAL collaboration for $W^{+}W^{-}\gamma\gamma$ couplings
are given by -0.020 $GeV^{-2} < \frac{a_{0}}{\Lambda^{2}} <$
0.020 $GeV^{-2}$, -0.052 $GeV^{-2} < \frac{a_{c}}{\Lambda^{2}} <$
0.037 $GeV^{-2}$ and for $ZZ\gamma\gamma$ couplings
-0.007 $GeV^{-2} < \frac{a_{0}}{\Lambda^{2}} <$
0.023 $GeV^{-2}$, -0.029 $GeV^{-2} < \frac{a_{c}}{\Lambda^{2}} <$
0.029 $GeV^{-2}$
at 95\% C.L. assuming that the $W^{+}W^{-}\gamma\gamma$
couplings are independent of the $ZZ\gamma\gamma$ couplings
\cite{opal}. 
If it is assumed that $W^{+}W^{-}\gamma\gamma$ couplings
are dependent on the $ZZ\gamma\gamma$ couplings (Same $a_{0}$ and $a_{c}$
parameters are used for both $W^{+}W^{-}\gamma\gamma$ and $ZZ\gamma\gamma$
couplings.) then the 95\% C.L. sensitivity limits are improved to
+0.002 $GeV^{-2} < \frac{a_{0}}{\Lambda^{2}} <$
0.019 $GeV^{-2}$, -0.022 $GeV^{-2} < \frac{a_{c}}{\Lambda^{2}} <$
0.029 $GeV^{-2}$.

Since the research and development on linear $e^{+}e^{-}$
colliders and its operating modes of $e^{+}e^{-}$, 
$e\gamma$ and $\gamma\gamma$ \cite{akerlof,barklow}
have been progressing there have  been several 
studies of anomalous quartic gauge boson couplings 
through the reactions $e^{+}e^{-} \to VVV$ \cite{barger},
 $e^{+}e^{-} \to FFVV$ \cite{boos},
$e\gamma \to VVF$ \cite{eboli} , 
$e\gamma \to VVVF$ \cite{eboli2}
$\gamma\gamma \to VV$ \cite{boudjema}, 
$\gamma\gamma \to VVV$ \cite{eboli3} and
$\gamma\gamma \to VVVV$ \cite{eboli2}
where $V=Z,W$ or $\gamma$ and $F=e$ or $\nu$.
These vertices have also been studied 
at hadron colliders via the process 
$pp(\bar{p})\to \gamma\gamma Z$ or $\gamma\gamma W$
\cite{eboli4}.

In this work  we consider the processes
$e\gamma \to \nu_{e} W \gamma$ and $e\gamma \to Z\gamma e$
to investigate $WW\gamma\gamma$ and 
$ZZ\gamma\gamma$ couplings. The bounds on these couplings 
were shown to be  weaker than the other anomalous quartic 
couplings discussed   in Ref.\cite{eboli} 
in $e\gamma$ collision. We take into account of the cross sections 
for longitudinal and transverse polarization states 
of the final W or Z boson as well as the incoming beam 
polarizations to improve the bounds, assuming 
the polarization of W and Z can be measured \cite{opal2}.

\section{Polarized real $\gamma$ beam}

Real gamma beam is obtained by the Compton
backscattering of laser photons off linear electron
beam where most of the photons are produced at the
high energy region. The luminosities for $e\gamma$
and $\gamma\gamma$ collisions turn out
to be of the same order as the one for $e^{+}e^{-}$
collision \cite{ginzb1}. This is the reason why one gets larger
cross section for photoproduction processes with real
photons.

The spectrum of backscattered photons is needed for
integrated cross section in connection
with helicities of initial laser photon and electron:

\begin{eqnarray}
f_{\gamma/e}(y)={{1}\over{g(\zeta)}}[1-y+{{1}\over{1-y}}
-{{4y}\over{\zeta(1-y)}}+{{4y^{2}}\over {\zeta^{2}(1-y)^{2}}}+
\lambda_{0}\lambda_{e} r\zeta (1-2r)(2-y)]
\end{eqnarray}
where

\begin{eqnarray}
g(\zeta)=&&g_{1}(\zeta)+
\lambda_{0}\lambda_{e}g_{2}(\zeta) \nonumber\\
g_{1}(\zeta)=&&(1-{{4}\over{\zeta}}
-{{8}\over{\zeta^{2}}})\ln{(\zeta+1)}
+{{1}\over{2}}+{{8}\over{\zeta}}-{{1}\over{2(\zeta+1)^{2}}} \\
g_{2}(\zeta)=&&(1+{{2}\over{\zeta}})\ln{(\zeta+1)}
-{{5}\over{2}}+{{1}\over{\zeta+1}}-{{1}\over{2(\zeta+1)^{2}}}
\end{eqnarray}
Here $r=y/[\zeta(1-y)]$, $y=E_{\gamma}/E_{e}$ and
$\zeta=4E_{e}E_{0}/M_{e}^{2}$. $E_{0}$ is the energy
of initial laser photon and $E_{e}$ and $\lambda_{e}$
are the energy and the helicity
of initial  electron beam before Compton backscattering. 
It is clear from $g(\zeta)$ spectrum $f(y)$ takes larger 
values when $\lambda_{0}\lambda_{e}$ has negative sign.
Because of this we keep this result during our 
computation. The maximum value of y reaches 0.83 when $\zeta=4.8$.

The helicity of the Compton backscattered photons can be written 
below in terms of the same parameters given above  

\begin{eqnarray}
\xi(E_{\gamma},\lambda_{0})={{\lambda_{0}(1-2r)
(1-y+1/(1-y))+\lambda_{e} r\zeta[1+(1-y)(1-2r)^{2}]}
\over{1-y+1/(1-y)-4r(1-r)-\lambda_{e}\lambda_{0}r\zeta
(2r-1)(2-y)}}.
\end{eqnarray}

which has the highest value when the parameter y is around 
its highest  value.

\section{Single W or Z Boson   Production in $e\gamma$ Collision}

The single production of W boson via process
$e\gamma\to W^{-}\gamma\nu_{e}$ is described by seven tree
level diagrams. Only t-channel W exchange diagram contains
anomalous $W^{+}W^{-}\gamma\gamma$ couplings.
The helicity amplitudes  have been calculated 
via vertex amplitude techniques \cite{maina} and
the phase space integrations have been performed 
by GRACE \cite{grace} which uses a Monte Carlo routine. 
In all calculations we impose a cut on the
final state photon transverse momentum $p_{T}^{\gamma}>15$ GeV.

In this work we assume that the transverse and the longitudinal
polarization states of the final W can be measured \cite{opal2}. 
This is a reasonable assumption since the 
angular distributions of the decay products of
the final W boson are closely related to the polarization states of it.
Therefore in principle, polarization states of the final W boson can
be determined by measuring the angular distributions of the W decay
products. Let us consider the differential cross section for the complete
process $e\gamma\to W^{-}\gamma\nu_{e} \to \ell\bar{\nu_{l}}\gamma\nu_{e}$
where lepton channel of the W boson decay is taken into account.
It can be written in the following form :

\begin{eqnarray}
d\sigma(e\gamma\to W^{-}\gamma\nu_{e} \to \ell\bar{\nu_{l}}\gamma\nu_{e})
=\frac {1}{32\pi M_{W} \Gamma_{W}}\sum_{\lambda_{W}} d\sigma_{a}
(\lambda_{W})|M_{b}(\lambda_{W})|^{2} d\cos{\theta}^{\star}
\end{eqnarray}
where $\lambda_{W}$  and $\Gamma_{W}$ show the polarization state
and the total decay width of W boson.
$d\sigma_{a}(\lambda_{W})$ is the helicity dependent differential
cross section for the process $e\gamma\to W^{-}\gamma\nu_{e}$ and
$M_{b}(\lambda_{W})$ is the
helicity amplitude for W decay to leptons ($W^{-} \to l\bar{\nu_{l}}$)
in the rest frame of W. ${\theta}^{\star}$ is the polar angle of the
final state leptons $l\bar{\nu_{l}}$ in the W rest frame with respect
to the W-boson direction in the $l\bar{\nu_{l}}\gamma\nu_{e}$ rest frame.

Decay amplitudes $|M_{b}(\lambda_{W})|^{2}$ have simple angular
dependences. When we take the helicity basis ($\lambda_{W}=+,-,0$)
explicit forms of these amplitudes in the W rest
frame are given by

\begin{eqnarray}
|M_{b}(+)|^{2}&&=\frac{g_{W}^{2}M_{W}^{2}}{4}
(1-\cos{\theta}^{\star} )^{2} \\
|M_{b}(-)|^{2}&&=\frac{g_{W}^{2}M_{W}^{2}}{4}
(1+\cos{\theta}^{\star})^{2} \\
|M_{b}(0)|^{2}&&=\frac{g_{W}^{2}M_{W}^{2}}{2}
(\sin{\theta}^{\star})^{2} \\
|M_{b}(LO)|^{2}&&=|M_{b}(0)|^{2} \\
|M_{b}(TR)|^{2}&&=|M_{b}(+)|^{2}+|M_{b}(-)|^{2}
\end{eqnarray}

\begin{figure}
\includegraphics{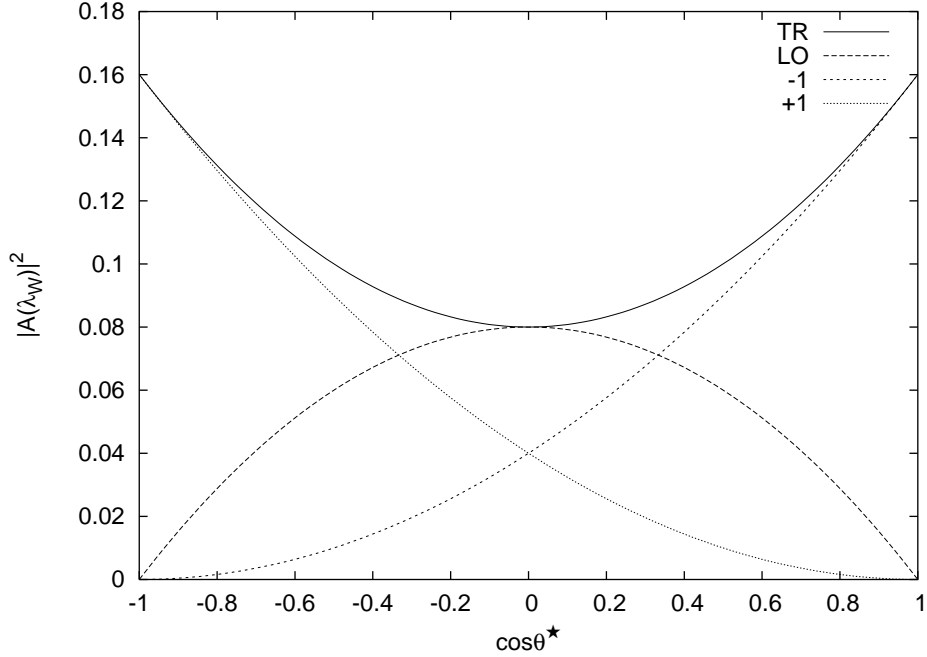}
\caption{Angular distribution of W decay products of
lepton channel in the rest system of the W boson where
$|A(\lambda_{W})|^{2}=\frac {1}{32\pi M_{W} \Gamma_{W}}
|M_{b}(\lambda_{W})|^{2}$. TR and LO stand for transverse
and longitudinal. 
\label{fig1}}
\end{figure}

By measuring the polar angle distributions of the W decay products,
one can directly determine the differential cross sections for fixed
W helicities. Complete factors $\frac {1}{32\pi M_{W} \Gamma_{W}}
|M_{b}(\lambda_{W})|^{2}$ in front of $d\sigma_{a}(\lambda_{W})$
in differential cross sections are plotted in Fig.\ref{fig1}. As
can be seen from Fig.~\ref{fig1} longitudinal(LO) and transverse(TR)
distributions are well separated from each other.

The single production of Z boson via process
$e\gamma\to Z\gamma e$ is also described by seven tree
level diagrams including only one  t-channel Z exchange 
diagram with anomalous $ZZ\gamma\gamma$ couplings.
The differential cross section for the complete process 
$e\gamma\to Z\gamma e \to \ell^{+}\ell^{-}\gamma e$
where lepton channel of the Z boson decay is chosen
can be written in a similar way to the W case above:

\begin{eqnarray}
d\sigma(e\gamma\to Z\gamma e \to \ell^{+}\ell^{-}\gamma e)
=\frac {1}{32\pi M_{Z} \Gamma_{Z}}\sum_{\lambda_{Z}} d\sigma_{a}
(\lambda_{Z})|M_{b}(\lambda_{Z})|^{2} d\cos{\theta}^{\star}
\end{eqnarray}

where $M_{b}(\lambda_{Z})$ is the helicity dependent amplitude 
for Z decay to leptons $Z\to \ell^{+}\ell^{-}$ in the rest 
frame of Z. $\theta^{\star}$ has the same definition as the 
case of W boson. Explicit form of the decay amplitudes 
$|M_{b}(\lambda_{Z})|^{2}$ with helicity basis 
$\lambda_{Z}=+,-,0$ in the Z rest frame are given as follows

\begin{eqnarray}
|M_{b}(+)|^{2}&&=\frac{M_{Z}^{2}}{2}
[g_{L}^{2}(1-\cos{\theta}^{\star} )^{2}+
g_{R}^{2}(1+\cos{\theta}^{\star} )^{2}  ] \\
|M_{b}(-)|^{2}&&=\frac{M_{Z}^{2}}{2}
[g_{L}^{2}(1+\cos{\theta}^{\star} )^{2}+
g_{R}^{2}(1-\cos{\theta}^{\star} )^{2}  ] \\
|M_{b}(0)|^{2}&&=M_{Z}^{2}(g_{L}^{2}+g_{R}^{2})
\sin^{2}{\theta}^{\star}
\end{eqnarray} 
Behaviour of $|M_{b}(TR)|^{2}$ and $|M_{b}(LO)|^{2}$ 
is almost the same as in the Fig.\ref{fig1}.

The  form of the helicity dependent differential 
cross section for $e\gamma\to Z\gamma e $  subprocess with initial beam 
polarizations is 

\begin{eqnarray}
{d\hat{\sigma}(\lambda_{0},\lambda_{Z})}=&&
 \{\frac{1}{4}(1-P_{e})
 \left[(1+\xi(E_{\gamma} , \lambda_{0}))|M(+,L;\lambda_{Z},L)|^{2}
+(1-\xi(E_{\gamma} , \lambda_{0}))|M(-,L;\lambda_{Z},L)|^{2}\right]
 \nonumber \\
+&&\frac{1}{4}(1+P_{e})
\left[(1+\xi(E_{\gamma} , \lambda_{0}))|M(+,R;\lambda_{Z},R)|^{2}
+(1-\xi(E_{\gamma} , \lambda_{0}))|M(-,R;\lambda_{Z},R)|^{2}\right]
\} \nonumber \\
&&\times (PS)
\end{eqnarray}
where helicity amplitudes 
$M(\lambda_{\gamma}, \sigma_{e}; \lambda_{Z}, \sigma_{e}^{\prime})$
are defined  to represent both incoming and
outgoing  electron helicities
$\sigma_{e}:L,R$  and $\sigma_{e}^{\prime}:L,R$  together with incoming and
outgoing  boson  helicities $\lambda_{\gamma}:+-$ , $\lambda_{Z}:+-0$.
The  polarization index of outgoing photon is not shown 
in the amplitude due to the sum  over its  polarization.    
Phase space and initial flux factors are included in  (PS).
Above cross section has been connected to initial laser photon helicity
$\lambda_{0}$ before Compton backscattering. $P_{e}$ is the initial 
electron beam polarization and is different from $\lambda_{e}$ which refers
to different beam. The information about the incoming photon beam
polarization $\xi(E_{\gamma}, \lambda_{0})$ (Compton backdcattered photons) 
can be found in previous section.
In the case of W production the last two terms beginning with $(1+P_{e})$
will  vanish.

The expression of the integrated cross section over
the backscattered photon spectrum is written below for
completeness:

\begin{eqnarray}
{d{\sigma}(\lambda_{0},\lambda_{Z})}
=\int_{y_{min}}^{y_{max}}
f_{\gamma/e}(y){d\hat{\sigma}(\lambda_{0},\lambda_{Z})}dy
\end{eqnarray}
with $y_{min}=M_{Z}^{2}/s$. Here $\hat{s}$ is related to $s$,
the square of the center of mass energy of $e^{+}e^{-}$ system,
by $\hat{s}=ys$.

\begin{figure}
\includegraphics{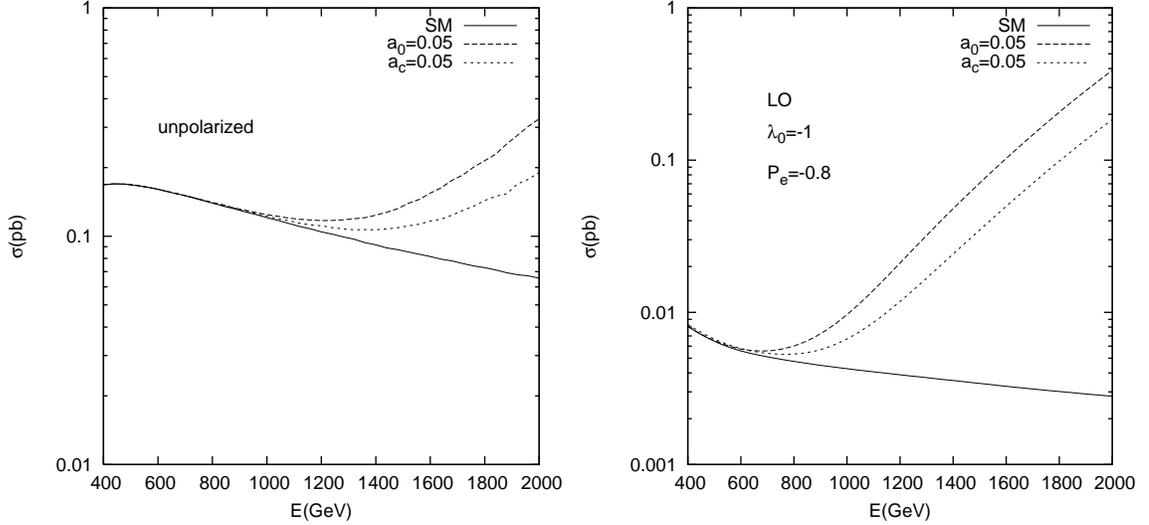}
\caption{The total cross sections of $e\gamma\to Z\gamma e$
as a function of collider energy for the SM and the anomalous 
couplings $a_{0}=0.05$, $a_{c}=0.05$ with unpolarized 
 and longitudinally polarized (LO)  final Z boson with polarized 
incoming beams. Only one of the anomalous couplings 
is kept different from their SM value.
\label{fig2}}
\end{figure}

In order to get an idea about the influence of the initial and final 
state polarizations on the cross section with anomalous 
$ZZ\gamma\gamma$ and $WW\gamma\gamma$ couplings
we give a set of figures. 
In all figures we take into account 
the polarization configuration which give the largest deviation 
from the Standard Model when compared to unpolarized cases. This
configuration corresponds to $\lambda_{0}=-1$, $P_{e}=-0.8$ 
and longitudinal polarization  of the final state Z or W boson. 
In all calculations $\lambda_{e}$ has been chosen to satisfy 
$\lambda_{0}\lambda_{e}<0$ to increase cross section. 
One can see from Fig.~\ref{fig2} the  deviations
of the total cross sections for the anomalous 
$ZZ\gamma\gamma$ couplings
$a_{0}=0.05$ and $a_{c}=0.05$ from their SM value
as a function of center of mass energy $\sqrt{s}$ of the
$e^{-}e^{+}$ system. Only one of the anomalous couplings
is kept different from the SM value. From these figures,
it is clear that the longitudinal polarization remarkably
improves the deviations from the SM at increasing energies
when compared to the unpolarized case. The transverse polarization
is ineffective on the variation of the cross section
for the given anomalous coupling values. Therefore, the
curves related to transverse polarization are not given.
The cross section of the process $e\gamma\to W^{-}\gamma\nu_{e}$
with anomalous $WW\gamma\gamma$ couplings shows the  similar 
features as a functon of energy. 

\begin{figure}
\includegraphics{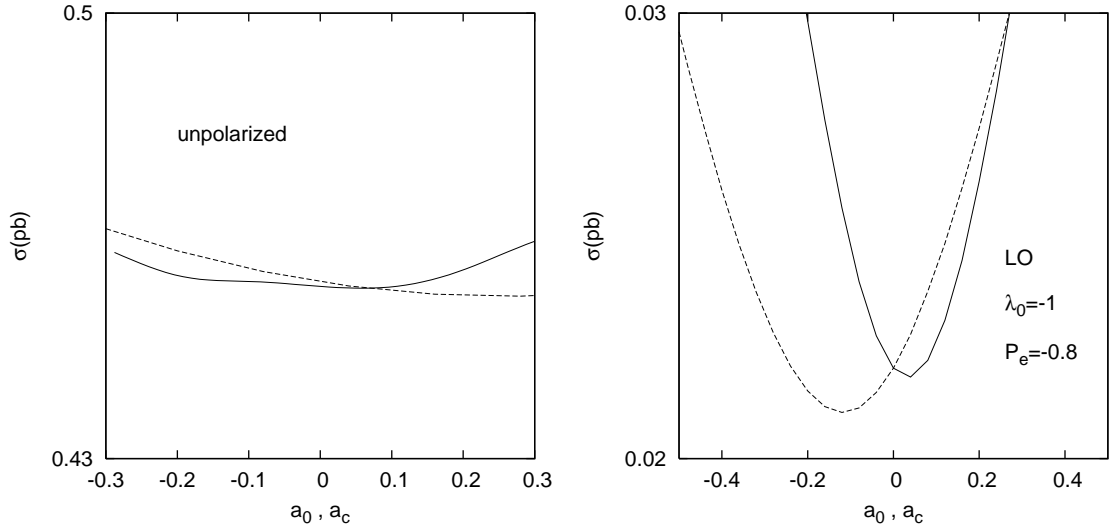}
\caption{Total cross sections of $e\gamma\to W^{-}\gamma\nu_{e}$
as a function of anomalous couplings
$a_{0}$ and $a_{c}$ for unpolarized and 
longitudinally polarized final W boson with incoming beam polarizations
at main  $e^{+}e^{-}$ collider energy $\sqrt{s}=0.5$ TeV.
Solid curves  show the  behaviour against the  coupling  $a_{0}$.
y axes are logarithmic scale.
\label{fig3}}
\end{figure}

\begin{figure}
\includegraphics{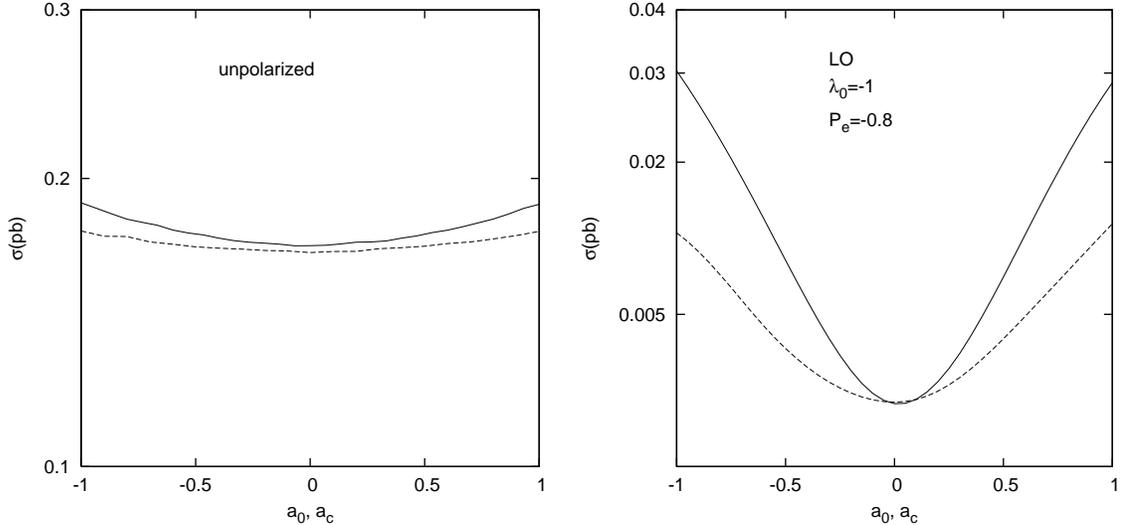}
\caption{Total cross sections of $e\gamma\to Z\gamma e$
as a function of anomalous couplings
$a_{0}$ and $a_{c}$ for unpolarized and
longitudinally polarized final Z boson with polarized 
incoming beams
at main  $e^{+}e^{-}$ collider energy $\sqrt{s}=0.5$ TeV.
Solid curves  show the  behaviour against the  coupling  $a_{0}$.
\label{fig4}}
\end{figure}

\begin{figure}
\includegraphics{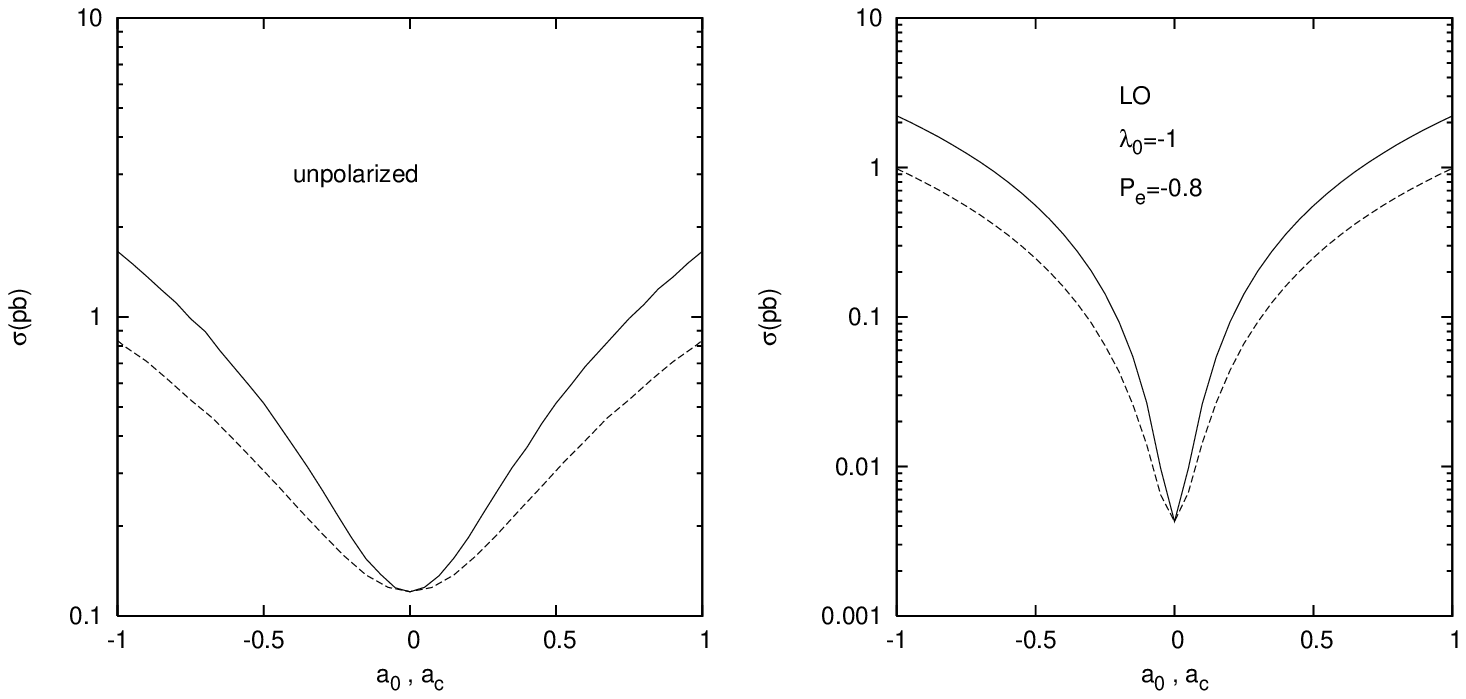}
\caption{The same as the Fig.~\ref{fig4} but for the energy 
$\sqrt{s}=1$ TeV.
\label{fig5}}
\end{figure}

In Fig.~\ref{fig3} and Fig.~\ref{fig4}
the total cross sections for $e\gamma\to W^{-}\gamma\nu_{e}$
and $e\gamma\to Z\gamma e$ as a function of anomalous couplings
$a_{0}$ and $a_{c}$ are plotted for above mentioned polarization 
configuration of the incoming beams and  final state 
W or Z bosons at the energy $\sqrt{s}=0.5$ TeV. Here, we assume that
the anomalous $W^{+}W^{-}\gamma\gamma$
couplings are independent of the $ZZ\gamma\gamma$ couplings.
The same curves are given in Fig.~\ref{fig5} for 
$e\gamma\to Z\gamma e$ at the collider energy $\sqrt{s}=1$ TeV.

From Fig.~\ref{fig3} and Fig.~\ref{fig4} it is clear that 
longitudinally polarized cross sections seem highly 
sensitive to anomalous couplings. Fig.~\ref{fig5} shows that anomalous 
couplings becomes far more effective as the energy increases.
In all figures we  realize the fact that $a_{0}$ dependence 
of the cross sections cause more separation from the SM. 
The comparison of the other polarization configurations to 
unpolarized cases and to each other will be given in the 
following section. 

\section{limits on the Anomalous Couplings}

For a concrete result,  95\% C.L. limits on the
anomalous coupling parameters
$a_{0}$ and $a_{c}$  have been obtained using
$\chi^{2}$ analysis at  $\sqrt{s}=0.5, 1$ TeV and
integrated luminosity $L_{int}=500$ $fb^{-1}$ without
systematic errors.
In order to get realistic results the number of events
are given as $N=AL_{int}\sigma BR $ where $A$ is the overall
acceptance and $BR$ is the branching ratio of W and Z boson
for leptonic channel.
The limits for the anomalous $WW\gamma\gamma$ couplings
are given on Table \ref{tab1} for unpolarized,
transverse and longitudinal polarization states of
final W boson and several combinations of incoming beams 
polarizations with the acceptance  $A=0.85$.
On Table \ref{tab2} and Table \ref{tab3}  the limits for the anomalous 
$ZZ\gamma\gamma$ couplings are given for more  
 polarization configurations based on the fact that 
fermion-fermion-Z coupling has both left and right handed 
parts.
One can see from Table \ref{tab1} that
longitudinal polarization together with the
initial beam polarizations improves limits at least
by a factor of 2-3 for  $a_{0}$ and a factor of 2 for $a_{c}$ at
$\sqrt{s}=0.5$ TeV. This improvement  reaches the limits
by a factor of 3-4  for $a_{0}$ but by a factor of 2-2.6 for $a_{c}$
at $\sqrt{s}=1$ TeV energy. Increase in energy from
0.5 TeV  to 1 TeV improves the limits  by a factor of 8 in the
longitudinally polarized and unpolarized case.
For the energies 1.5 TeV and higher, the improvement due to
longitudinal polarization is expected to be  much  better
than the unpolarized case. Table \ref{tab2} and Table \ref{tab3} show
 similar features but with slightly worse limits, 2-2.6 
depending on the polarizations and energy for $ZZ\gamma\gamma$ couplings. 
From these three tables, we realize 
that the improvement factor caused by initial beam polarizations alone 
is at most 1.3. 

For $ZZ\gamma\gamma$ couplings the best limit comes from the 
process $\gamma\gamma\to ZZ$ due to absence of tree level 
SM background \cite{boudjema}. The bounds on the anomalous 
$WW\gamma\gamma$ couplings has the same order as 
the ones from the process $\gamma\gamma\to WW$ \cite{boudjema}.
As can be seen from the references given in the introduction, 
the limits obtained in this work for the longitudinal cases
are better than the other 
basic leptonic and hadronic colliders such as the linear 
$e^{+}e^{-}$ collider and CERN LHC.   

\begin{table}
\caption{Sensitivity of the  $e\gamma $ collision
to $WW\gamma\gamma$ couplings at 95\% C.L. for
$\sqrt{s}=0.5, 1$ TeV and $L_{int}=500$ $fb^{-1}$. The effects of
final state W boson polarization and initial beam 
polarizations are shown in each row. Here 
$\lambda_{0}$ is the laser photon polarization before 
Compton backscattering and  $P_e$ electron beam polarization.
Only one of the couplings is assumed to deviate
from the SM at a time. LO and TR stand for longitudinal and 
transverse.
\label{tab1}}
\begin{ruledtabular}
\begin{tabular}{cccccc}
$\sqrt{s}$ TeV & $\lambda_{0}$ & $P_{e}$  &  $\lambda_{W}$ 
& $a_{0}$ & $a_{c}$  \\
\hline
0.5 & 0 &0 & TR+LO & -0.26, 0.23
& -0.20, 0.69  \\
0.5 & 0 &0 & LO & -0.16, 0.07
& -0.08, 0.32  \\
0.5 & 0 &0 & TR & -0.49, 1.30
& -0.3, 0.4 and 2.6, 3.2  \\
0.5 & 1 &-0.8 & TR+LO & -0.20, 0.20
& -0.13, 0.60  \\
0.5 & -1 &-0.8 & TR+LO & -0.17, 0.18 &
-0.10, 0.21 and 0.29, 0.65  \\
0.5 & 1 &-0.8 & LO & -0.10, 0.06
& -0.06, 0.27  \\
0.5 & 1 &-0.8 & TR & -0.5, 1.25
& -0.29, 0.38 and 1.40, 2.00   \\
0.5 & -1 &-0.8 & LO & -0.12, 0.05
& -0.05, 0.29  \\
0.5 & -1 &-0.8 & TR & -0.66, 1.20
& -0.15, 0.25 and 2.45, 3.00  \\
\hline
1 & 0 &0 & TR+LO & -0.04, 0.04
& -0.03, 0.07  \\
1 & 0 &0 & LO & -0.02, 0.01
& -0.01, 0.04  \\
1 & 0 &0 & TR & -0.17, 0.34
& -0.10, 0.61  \\
1 & 1 &-0.8 & TR+LO & -0.03, 0.03
& -0.03, 0.04  \\
1 & -1 &-0.8 & TR+LO & -0.03, 0.03
& -0.03, 0.07  \\
1 & 1 &-0.8 & LO & -0.01, 0.01
& -0.01, 0.03  \\
1 & 1 &-0.8 & TR & -0.20, 0.35
& -0.09, 040.  \\
1 & -1 &-0.8 & LO & -0.01, 0.01
& -0.01, 0.03  \\
1 & -1 &-0.8 & TR & -0.34, 0.33
& -0.05, 0.5  
\end{tabular}
\end{ruledtabular}
\end{table}

\begin{table}
\caption{Sensitivity of the  $e\gamma $ collision
to $ZZ\gamma\gamma$ couplings at 95\% C.L. for
$\sqrt{s}=0.5$ TeV and $L_{int}=500$ $fb^{-1}$. The effects of
final state Z boson polarization and  initial beam 
polarizations are shown in each row.
Only one of the couplings is assumed to deviate
from the SM at a time.\label{tab2}}
\begin{ruledtabular}
\begin{tabular}{cccccc}
$\sqrt{s}$ TeV & $\lambda_{0}$ & $P_{e}$  &  $\lambda_{Z}$ 
& $a_{0}$ & $a_{c}$  \\
\hline
0.5 & 0 &0 & TR+LO & -0.50, 0.50
& -0.70, 0.70  \\
0.5 & 0 &0 & LO & -0.20, 0.30
& -0.40, 0.40  \\
0.5 & 0 &0 & TR & -1.70, 1.20
& -1.80, 1.80  \\
0.5 & 1 &-0.8 & TR+LO & -0.44, 0.42
& -0.56, 0.64  \\
0.5 & -1 &-0.8 & TR+LO & -0.37, 0.36 &
-0.54, 0.60   \\
0.5 & 1 &0.8 & TR+LO & -0.43, 0.43
& -0.64, 0.65  \\
0.5 & -1 &0.8 & TR+LO & -0.45, 0.45 &
-0.60, 0.65   \\
0.5 & 1 &-0.8 & LO & -0.21, 0.24
& -0.34, 0.30  \\
0.5 & 1 &-0.8 & TR & -1.60, 1.42
& -1.00, 1.70    \\
0.5 & -1 &-0.8 & LO & -0.19, 0.24
& -0.31, 0.29  \\
0.5 & -1 &-0.8 & TR & -1.60, 1.21
& -1.25, 1.50  \\
0.5 & 1 &0.8 & LO & -0.21, 0.25
& -0.34, 0.32  \\
0.5 & 1 &0.8 & TR & -1.75, 1.45
& -1.35, 1.75    \\
0.5 & -1 &0.8 & LO & -0.23, 0.27
& -0.36, 0.32  \\
0.5 & -1 &0.8 & TR & -1.79, 1.52
& -1.29, 1.87    \\
\end{tabular}
\end{ruledtabular}
\end{table}

\begin{table}
\caption{Sensitivity of the  $e\gamma $ collision
to $ZZ\gamma\gamma$ couplings at 95\% C.L. for
$\sqrt{s}=1$ TeV and $L_{int}=500$ $fb^{-1}$. The effects of
final state Z boson polarization and  initial beam
polarizations are shown in each row.
Only one of the couplings is assumed to deviate
from the SM at a time.\label{tab3}}
\begin{ruledtabular}
\begin{tabular}{cccccc}
$\sqrt{s}$ TeV & $\lambda_{0}$ & $P_{e}$  &  $\lambda_{Z}$
& $a_{0}$ & $a_{c}$  \\
\hline
1 & 0 &0 & TR+LO & -0.05, 0.05
& -0.07, 0.07  \\
1 & 0 &0 & LO & -0.02, 0.02
& -0.03, 0.03  \\
1 & 0 &0 & TR & -0.30, 0.30
& -0.30, 0.30  \\
1 & 1 &-0.8 & TR+LO & -0.04, 0.04
& -0.05, 0.05  \\
1 & -1 &-0.8 & TR+LO & -0.04, 0.04
& -0.06, 0.06  \\
1 & 1 &0.8 & TR+LO & -0.04, 0.04
& -0.06, 0.06  \\
1 & -1 &0.8 & TR+LO & -0.04, 0.04
& -0.06, 0.06  \\
1 & 1 &-0.8 & LO & -0.02, 0.02
& -0.03, 0.03  \\
1 & 1 &-0.8 & TR & -0.37, 0.34
& -0.23, 0.26  \\
1 & -1 &-0.8 & LO & -0.02, 0.02
& -0.03, 0.03  \\
1 & -1 &-0.8 & TR & -0.37, 0.34
& -0.26, 0.29  \\
1 & 1 &0.8 & LO & -0.02, 0.02
& -0.03, 0.03  \\
1 & 1 & 0.8 & TR & -0.40, 0.39
& -0.27, 0.30 \\
1 & -1 &0.8 & LO & -0.02, 0.02
& -0.03, 0.03  \\
1 & -1 & 0.8 & TR & -0.37, 0.34
& -0.24, 0.29 \\
\end{tabular}
\end{ruledtabular}
\end{table}

\pagebreak

\end{document}